\documentclass{ws-ijmpcs}

\newcommand{\mcdot}{\!\cdot\!}

\newcommand{\e}{\mathrm{e}}

\newcommand{\eq}[1]{Eq.~\eqref{#1}}

\def\SCETG{${\rm SCET}_{\rm G}\,$}

\newcommand{\be}{\begin{equation}}
\newcommand{\ee}{\end{equation}}

\def\d{{\rm d}}
\def\OMIT#1{{}}

\newcommand{\vc}[1]{{\bf{#1}}}

\def\SCETG{${\rm SCET}_{\rm G}\,$}

\def\trimmarks{}

\begin{document}
\markboth{G.Ovanesyan}
{Jet quenching beyond the energy loss approach}

%
\catchline{}{}{}{}{}
%

\title{JET QUENCHING BEYOND THE ENERGY LOSS APPROACH
}

\author{{{GRIGORY OVANESYAN}}
}

\address{Physics Department, University of Massachusetts Amherst\\
 Amherst, MA 01003, USA
\\
ovanesyan@umass.edu}



\maketitle

\begin{history}
\end{history}

\begin{abstract}
We study the jet quenching effect in heavy ion collisions, based on medium-induced splitting functions calculated from Soft Collinear Effective Theory with Glauber Gluons. Our method is formulated in the language of DGLAP evolution equations with medium-induced splitting functions. In the small-$x$ soft gluon approximation we analytically solve the evolution equations and find an intuitive connection to the energy loss approach. For central Pb+Pb collisions at the LHC we quantify the effect of finite-$x$ corrections for the nuclear modification factor and compare to data.
\end{abstract}

\ccode{PACS numbers:  25.75.-q, 12.38.Bx, 12.38.Cy}
\ccode{Preprint number: ACFI-T14-20}

\section{Introduction}	

Jet quenching [\refcite{Wang:1991xy,Vitev}] is a powerful probe of the properties of the quark-gluon plasma (QGP) created in heavy ion collisions. A perturbative QCD approach based on energy-loss calculations [\refcite{Gyulassy:2003mc}] has been successful in describing the jet quenching data in heavy ion collisions. One approximation that is inherent in energy-loss based predictions is that the fractional energy loss for the partons is small, i.e. the value $x=E/E_0\ll 1$, where $E_0$ is the energy of the parent parton and $E$ is the energy of the emitted parton in the $1\rightarrow 2$ splitting in the radiative process. We will refer to this approximation as either small-$x$ (soft gluon) approximation or energy loss approximation.

In the past several years progress has been made on extending the perturbative QCD calculations of medium-induced processes beyond the small-$x$ approximation. This became possible with the  formulation of an effective theory for jets in the dense QCD matter. Typically, in perturbative calculations the medium is modeled as consisting of effective scattering centers that act as sources of colored Coulomb potential~[\refcite{Gyulassy_WANG_Model}]. Therefore, a successful effective theory must describe the usual vacuum interactions of boosted partons with small invariant mass, as well as elastic scattering of energetic partons off of the medium quasiparticles. The former interactions are described by the familiar Soft Collinear Effective Theory (SCET) [\refcite{SCET1}--\refcite{SCET4}], and the latter ones need to be added to SCET. The resulting Soft Collinear Effective Theory with Glauber gluons (\SCETG) was formulated in Refs.[\refcite{SCETG1}--\refcite{SCETG3}]. Besides being a useful tool for heavy ion collisions, it is known that \SCETG has important applications for the Drell-Yan process~[\refcite{Bauer:2010cc}] and for the Regge physics~[\refcite{Donoghue:2009cq}--\refcite{Donoghue:2014mpa}]. More work is anticipated in these directions.

Using \SCETG, all four medium-induced splitting kernels have been calculated beyond the small-$x$ approximation in Ref.[\refcite{Ovanesyan:2011kn}]. However, it is not possible to consistently implement these results into the energy-loss formalism. This is because the energy-loss interpretation is only valid in the soft gluon emission limit. In a more recent work in Ref~[\refcite{Kang:2014xsa}] we formulated the nuclear modification factor in terms of the medium-modified Fragmentation Functions (FFs), which in turn are found via Dokshitzer-Altarelli-Parisi-Gribov-Lipatov (DGLAP) evolution equations~[\refcite{DGLAP1}--\refcite{DGLAP3}]. In these evolution equations we use the splitting functions based on the \SCETG calculations [\refcite{Ovanesyan:2011kn}] of all medium-induced splitting kernels, valid beyond the small-$x$ approximation. This allowed us to consistently incorporate and study the jet quenching  beyond the energy loss approach. In Refs.~[\refcite{Wang:2009qb,Chang:2014fba}] the hadron production in deep inelastic scattering has been studied from DGLAP evolution equations in the medium with initial conditions obtained using the energy loss approach. In the rest of this talk we concentrate on presenting our results from Ref.~[\refcite{Kang:2014xsa}]. 

\section{Medium-induced splitting functions}
Let us recall how the vacuum splitting functions in QCD are obtained~[\refcite{DGLAP3}]. First the real emission graphs need to be evaluated. To leading order in the strong coupling they are given by tree level Feynman graphs for the $q\rightarrow qg, g\rightarrow gg, g\rightarrow q\bar{q}, q\rightarrow gq$ splittings. Next, the virtual graphs need to be evaluated which contribute only for the first two splittings. Alternatively one can take advantage of momentum and flavor sum rules~[\refcite{DGLAP3}] and uniquely determine all the virtual pieces. As a result the first two splitting functions have a plus function ($1/x_+$) and a delta function pieces ($\delta(x)$), while the remaining two splittings are not singular as $x\rightarrow 0$ and do not need such a prescription.

The real emission graphs of the medium-induced splitting functions have been calculated in Ref.[\refcite{Ovanesyan:2011kn}] to  first order in the opacity expansion  [\refcite{Opacity_Expansion_1}, \refcite{Opacity_Expansion_2}]. In the presence of dense QCD matter the splitting functions are defined in the following~way\footnote{Note, that in Ref.[\refcite{Ovanesyan:2011kn}] we evaluated the medium-induced splitting kernels $x\d N/\d x$ which are related to the medium-induced splitting functions as $P^{\text{real}}_{i}(x,\vc{Q}_{\perp};\beta)=\frac{2\pi^2}{\alpha_s}\,\vc{Q}_{\perp}^2\,\frac{\d N_i (x, \vc{Q}_{\perp};\beta)}{\d x\,\d^2\vc{Q}_{\perp}} .$ }: 
\begin{eqnarray}
&&\left\langle \left|\mathcal{M}^{(1)}_{n+1,i}\right|^2+2\text{Re}\left(\mathcal{M}^{(0)}_{n+1,i}\right)^*\mathcal{M}^{(2c)}_{n+1,i} \right\rangle_{\vc{q}_{\perp}}=\frac{2g^2}{p_n^2} P^{\text{real}}_{i}(x,\vc{Q}_{\perp};\beta)\left\langle\left|\mathcal{M}^{(0)}_{n}\right|^2\right\rangle ,  \label{eq:Pdef1}
\end{eqnarray}
where $\mathcal{M}^{(0)}_{n+1,i}, \mathcal{M}^{(1)}_{n+1,i}, \mathcal{M}^{(2c)}_{n+1,i}$ are the amplitudes in which one of the hard partons in the process splits into two partons ($n\rightarrow n+1$) for the vacuum, first order Born graphs and contact limit of the second order Born graphs correspondingly, $\mathcal{M}^{(0)}_{n}$ is the production amplitude of the hard process before the splitting has occured. The splitting type is described by the subscript $i$ and is one of the four types mentioned above. On the right-hand side of the equation above we made explicit the dependence of the medium-induced splitting function on the fraction of energy carried by the emitted parton $x$, its transverse momentum $\vc{Q}_{\perp}$ with respect to the decaying parton $n$, and the properties of the medium $\beta$. Using explicit formulas based on \SCETG calculations of medium-induced kernels in Ref.[\refcite{Ovanesyan:2011kn}] we get
\begin{eqnarray}
P^{\text{real}}_{i}(x,\vc{Q}_{\perp};\beta)= P^{\text{vac}}_i(x)\,h_i(x,\vc{Q}_{\perp};\beta)\,,\label{eq:Pdef2}
\end{eqnarray}
where $P^{\text{vac}}_i(x)$ are the real emission parts of the vacuum splitting functions~[\refcite{DGLAP3}] and the splitting-type dependent functions $h_i$ are listed in the \ref{sec:appendix}. Proceeding by analogy with the vacuum calculation of virtual pieces of splitting functions, we use the flavor and momentum sum rules and find the virtual pieces of the medium-induced splitting functions. For the full splitting functions we get~[\refcite{Kang:2014xsa}]
\begin{eqnarray}
&&P_{q\rightarrow qg}(x,\vc{Q}_{\perp};\beta)=\left[P^{\text{real}}_{q\rightarrow qg}(x,\vc{Q}_{\perp};\beta)\right]_++A(\vc{Q}_{\perp};\beta)\,\delta(x)\, ,\nonumber\\
&&P_{g\rightarrow gg}(x,\vc{Q}_{\perp};\beta)=2C_A\Bigg\{\left[\left(\frac{1-2x}{x}+x(1-x)\right)h_{g\rightarrow gg}(x,\vc{Q}_{\perp};\beta)\right]_+\nonumber\\
&&\,\,\,\qquad\qquad\qquad\qquad\qquad+\frac{h_{g\rightarrow gg}(x,\vc{Q}_{\perp};\beta)}{1-x}\Bigg\}+B(\vc{Q}_{\perp};\beta)\,\delta(x)\, ,\nonumber\\
&&P_{g\rightarrow q\bar{q}}(x,\vc{Q}_{\perp};\beta)=P^{\text{real}}_{g\rightarrow q\bar{q}}(x,\vc{Q}_{\perp};\beta)\, ,\nonumber\\
 &&P_{q\rightarrow gq}(x,\vc{Q}_{\perp};\beta)=P^{\text{real}}_{q\rightarrow gq}(x,\vc{Q}_{\perp};\beta)\, ,\label{eq:fullsplittingfunctions}
\end{eqnarray}
where the plus function is with respect to the $x$ variable at $x=0$. More precisely, in the top two equations of \eq{eq:fullsplittingfunctions} the plus function is associated with the entire square bracket. For example, in the first line, defining $P^{\text{real}}_{q\rightarrow qg}(x,\vc{Q}_{\perp};\beta)\approx a/x$ as $x\rightarrow 0$, we have $[a/x]_+$ and a virtual piece $A$. However, if we rewrite this function in terms of $[1/x]_+$ we would get a virtual piece different from $A$. Note that the apparent singularity of the $g\rightarrow gg$ splitting as $x\rightarrow 1$ need not be regulated by the plus function, as it is regulated by the form of the evolution equations, similarly to the vacuum case for this splitting. The functions $A,B$ are equal to~[\refcite{Kang:2014xsa}]
\begin{eqnarray}
&&A(\vc{Q}_{\perp};\beta) = 0\, , \nonumber\\
&&B(\vc{Q}_{\perp};\beta) =\int_0^1\d x'\Bigg\{-2n_f(1-x')P_{g\rightarrow q\bar{q}}(x',\vc{Q}_{\perp};\beta)\nonumber\\
&&+2C_A\Bigg[x'\left(\frac{1-2x'}{x'}+x'(1-x')\right)-1\Bigg]h_{g\rightarrow gg}\left(x',\vc{Q}_{\perp};\beta\right)\Bigg\}\, . \quad \label{eq:ABvalues}
\end{eqnarray}
\section{Evolution equations and jet quenching}

Having determined the full medium-induced splitting functions, including the virtual pieces, we move on to study the evolution of FFs in the medium. We use the DGLAP evolution equations
\begin{eqnarray}
&&\frac{\d D_q(z,Q;\beta)}{\d \ln Q}=\frac{\alpha_s(Q^2)}{\pi}\int_{z}^1 \frac{\d z'}{z'}\Big[P_{q\rightarrow qg}(z',Q;\beta)D_q\left(\frac{z}{z'},Q;\beta\right)\nonumber\\
&&\qquad\qquad\qquad\qquad+P_{q\rightarrow gq}(z',Q;\beta)D_g\left(\frac{z}{z'},Q;\beta\right)\Big]\, ,\nonumber\\
&&\frac{\d D_g(z,Q;\beta)}{\d \ln Q}=\frac{\alpha_s(Q^2)}{\pi}\int_{z}^1 \frac{\d z'}{z'}\Bigg[P_{g\rightarrow gg}(z',Q;\beta)D_g\left(\frac{z}{z'},Q;\beta\right)\nonumber\\
&&\qquad\qquad\qquad\qquad+P_{g\rightarrow q\bar{q}}(z',Q;\beta)\sum_{q}\left(D_q\left(\frac{z}{z'},Q;\beta\right)+D_{\bar{q}}\left(\frac{z}{z'},Q;\beta\right)\right)\Bigg]\, .\label{eq:AP30}
\end{eqnarray}
First, note that in the equations above we have switched from the $x$ variable to the $z=1-x$ ($z'=1-x'$). Second, the form of evolution equations above is identical to the traditional DGLAP evolution equations in the vacuum, except one qualitative difference, that the medium-induced splitting functions also depend on $Q=|\vc{Q}_{\perp}|$ (and medium properties $\beta$), while the vacuum splitting functions only depend on $z$. The anti-quark FF evolution equation is same as for the quark up to $D_q\rightarrow D_{\bar{q}}\,$ on both sides. Because the QGP-induced processes are 
final-state effects in the heavy ion collisions, one can use the vacuum unmodified parton distribution functions (PDFs), and only the FFs are subject to medium-effects. 
\begin{figure}[!t]
\begin{center}
\includegraphics[width=7cm]{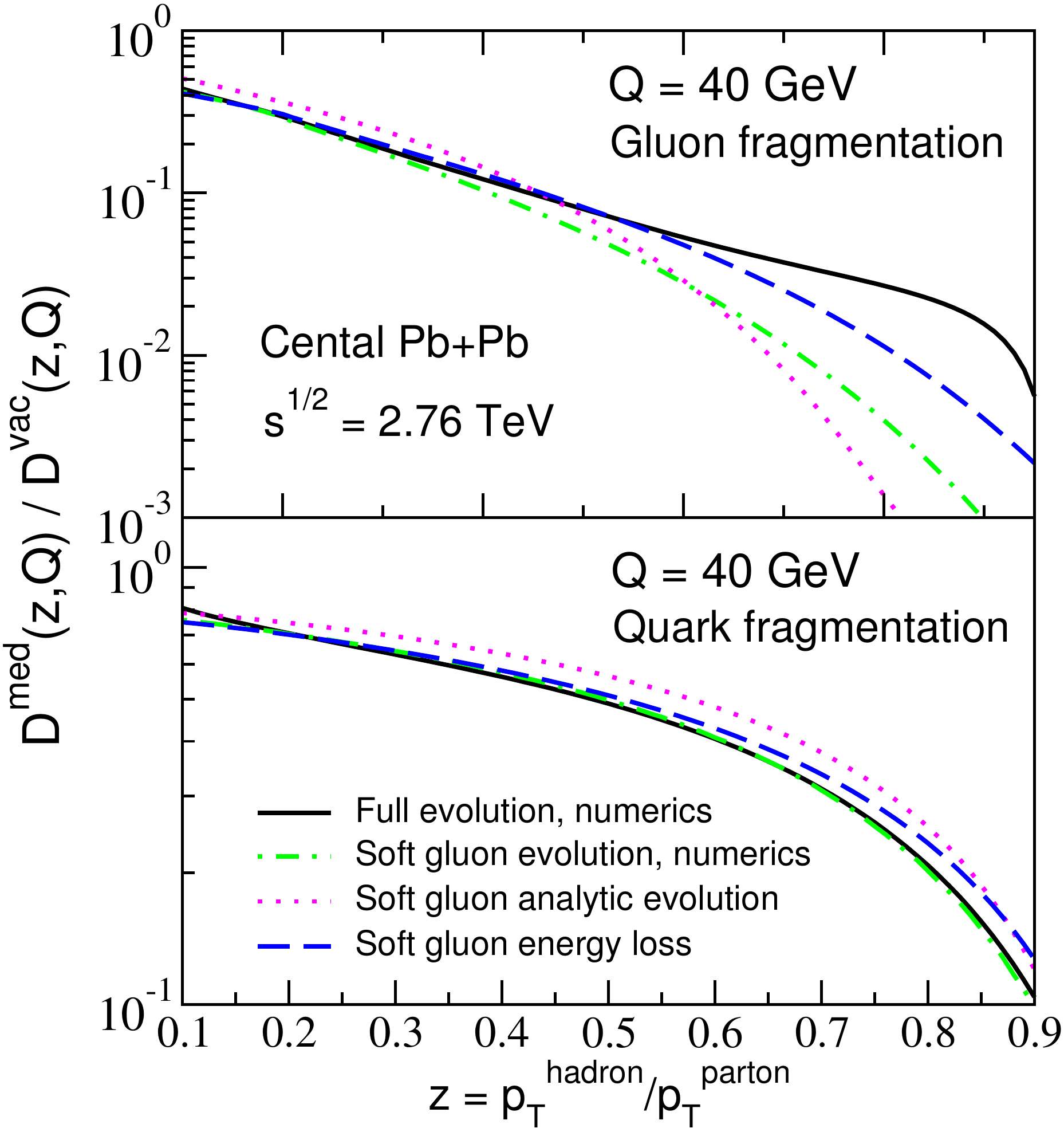}
\end{center}\caption{Modification of the medium-induced FF $D^{\text{med}}/D^{\text{vac}}$. We consider the most central Pb+Pb collisions at the LHC at $\sqrt{s}=2.76$\,TeV and a quark and a gluon jet with $p_{T_c}=Q=40\text{ GeV}$. We present exact solution to DGLAP evolution equations, small-$x$ (soft gluon) approximation numerical and analytical solutions, and the energy loss approach. }\label{fig:ffevolution}
\end{figure}

For the special case of the soft gluon approximation, the only two splittings that are relevant are $q\rightarrow qg$ and $g\rightarrow gg$ and the splitting functions become purely plus functions. Also, in the evolution equations in this approximation there is decoupling between the FFs, and each of them evolves independently from the others. This evolution equation for either of the FFs becomes
\begin{eqnarray}
\frac{\d D(z,Q)}{\d\ln Q}=\frac{\alpha_s}{\pi}\int_z^{1}\frac{\d z'}{z'}\,\left[P(z',Q)\right]_+D\left(\frac{z}{z'},Q\right) \,. \qquad
\label{eq:evolutionfull}
\end{eqnarray}
We define the steepness of unmodified vacuum FFs $n(z)=-{\d \ln D^{\text{vac}}(z)}/{\d\ln z}\,$\,. We use the approximation that the main support from the convolution integral in the evolution equation comes from the neighborhood of $z'\approx 1$. In this limit, the logarithmic derivative of the fragmentation function appears on the right hand side of \eq{eq:evolutionfull}, and by using the unperturbed vacuum value for this derivative, we find an analytical approximate formula~[\refcite{Kang:2014xsa}]
\begin{eqnarray}
&&D^{\text{med}}(z,Q)\approx \e^{-(n(z)-1)\left\langle\frac{\Delta E}{E}\right\rangle_z-\left\langle N_g\right\rangle_z}D^{\rm vac}(z,Q)\,, \qquad\label{eq:masterconnect}
\end{eqnarray}
where
\begin{eqnarray}
&&\left\langle\frac{\Delta E}{E}\right\rangle_z=\int_0^{1-z}\d x \,x\,\frac{\d N}{\d x}(x) \xrightarrow{z \to 0}\left\langle\frac{\Delta E}{E}\right\rangle, \,\,\, \left\langle N_g\right\rangle_z=\int_{1-z}^1\d x\,\frac{\d N}{\d x}(x)\xrightarrow{z\to1}\left\langle N_g\right\rangle\,.\nonumber
\end{eqnarray}
Our approximate method shows explicitly how the medium and vacuum radiative effects factorize, which we emphasized in \eq{eq:masterconnect}. The quantities $\left\langle\frac{\Delta E}{E}\right\rangle_z$ and $\left\langle N_g\right\rangle_z$ also depend on $Q$ via the upper limit of integration in $\d N/\d x=\int_{Q_0}^{Q} \d N/(\d x\,\d Q')\,\d Q'$\,.

The analytical formula obtained in \eq{eq:masterconnect} gives us an insight on the connection between the evolution method in the small-$x$ approximation and the energy loss approach. We see that for small to intermediate values of $z$ the entire quenching factor is given by an exponential $R_{AA}\sim\e^{-n\langle\Delta E/E\rangle}$ ($n\gg 1$), i.e. controlled by the steepness of the FFs times the fractional energy loss, while for the values of $z\approx 1$ we get $R_{AA}\sim\e^{-\langle N_g\rangle}$, which is the probability of non-branching. This behavior is the same in the energy loss approach and is an encouraging feature that the two approaches behave qualitatively the same way.

In terms of the evolved FFs the nuclear modification factor equals
\begin{eqnarray}
R^h_{AA}(p_T)=\frac{\sum_{c=q,\bar{q},g}\int^1_{z_{\text{min}}}\d z\frac{\d\sigma^c(p_{T_c}=p_T/z)}{\d y \d^2 p_{{T}_c}}\frac{1}{z^2}D_{c\rightarrow h}^{\text{med/quench}}(z,p_{T_c};\beta)}{\sum_{c=q,\bar{q},g}\int^1_{z_{\text{min}}}\d z\frac{\d\sigma^c(p_{T_c}=p_T/z)}{\d y \d^2 p_{{T}_c}}\frac{1}{z^2}D_{c\rightarrow h}^{\text{vac}}(z,p_{T_c})}\,,
\end{eqnarray}
where we have chosen $\mu=Q=p_{T_c}$ ($\mu$-the factorization scale) and therefore unmodified hard production partonic cross sections are used $\frac{\d\sigma^c(p_{T_c}=p_T/z)}{\d y \d^2 p_{{T}_c}}$ {\em per binary nucleon-nucleon collision}.

In Figure~\ref{fig:ffevolution} we compare the medium-induced FFs found from the evolution equations with that from the energy loss approach. We plot quark and gluon modified FFs for a $p_{T_c}=Q=40$ GeV jet created in $\sqrt{s}=2.76$ TeV Pb+Pb collisions at the LHC. We show the results for full-$x$ evolution (solid black lines), small $x$ evolution numerical (dot-dashed green lines), small-$x$ evolution analytical (dotted red lines) and for the energy loss (long-dashed blue lines) approaches. Note that, for both quark and gluon cases, the analytical and numerical small-$x$ evolution lines agree well for the most values of $z$. For the quark fragmentation function all the lines including the full $x$ numerical and the energy loss results are close to each other. For the gluon case the differences are more visible especially for $z>0.6$. However the nuclear modification factor is more sensitive to the $D_q^{\text{med}}$, because $D_g^{\text{med}}$ is more quenched due to a larger casimir factor $C_A>C_F$.
\begin{figure}[!t]
\includegraphics[width=6.3cm]{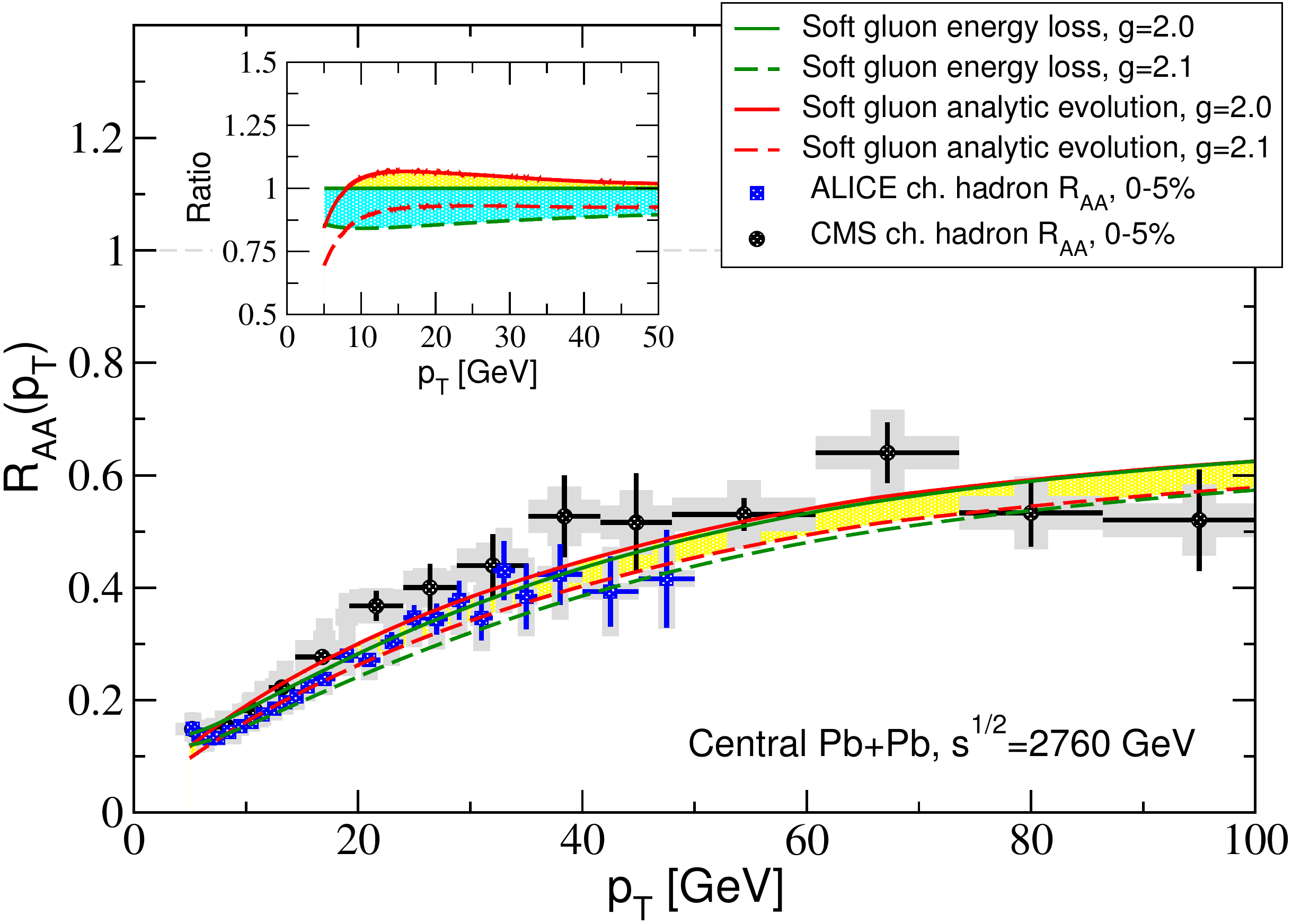}\,\,\,\includegraphics[width=6.3cm]{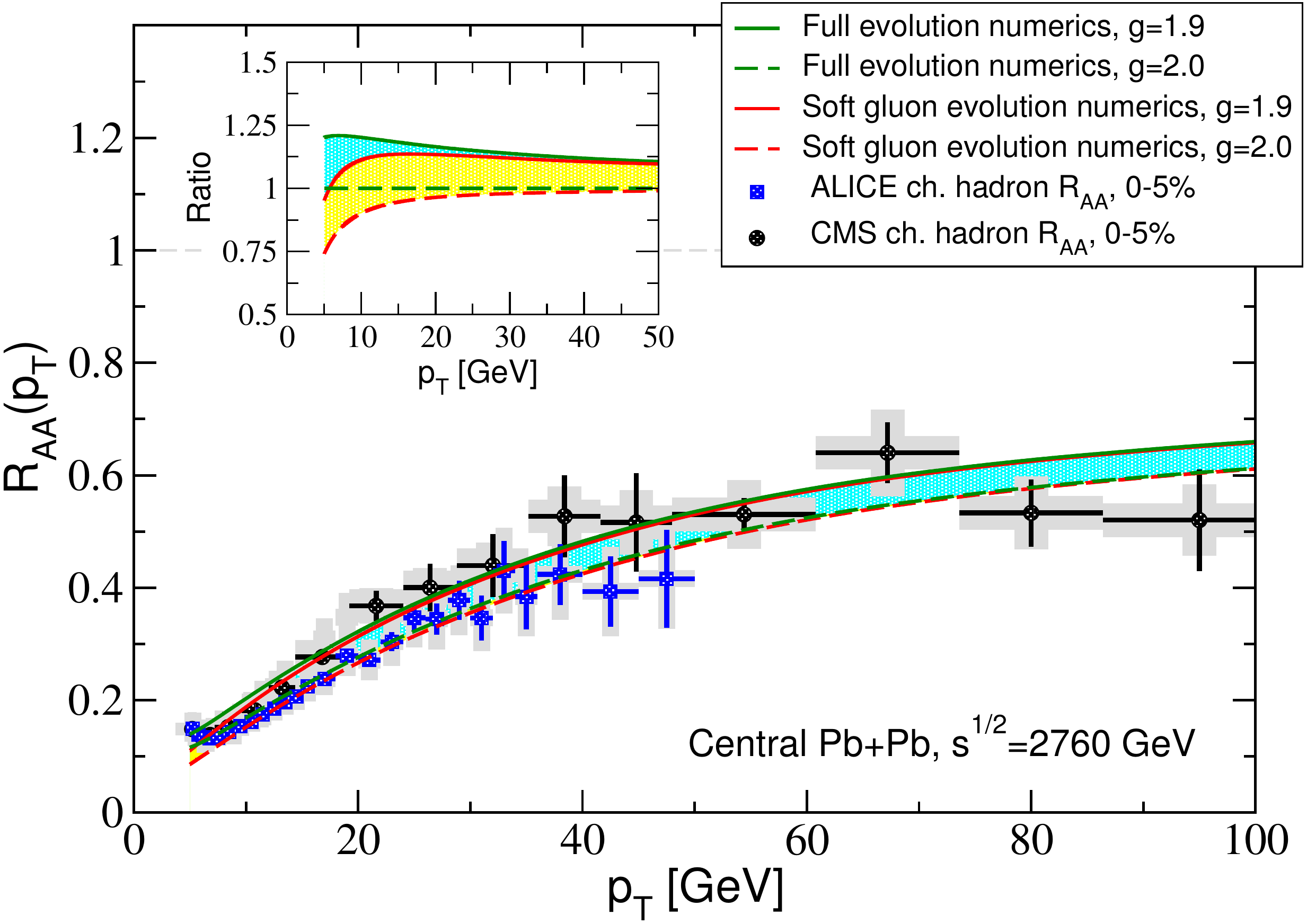}\caption{Nuclear modification factor $R_{AA}$ in the central Pb+Pb collisions for $\sqrt{s}=2.76$\,TeV energy. Left: comparison of the small-$x$ (soft gluon) evolution approach to the energy loss approach. Right: comparison of 
full-$x$ to the small-$x$ evolution approaches. Data from LHC included.}\label{fig:jetquenching}
\end{figure}

In Figure~\ref{fig:jetquenching} we show the comparison of the predictions based on different theoretical approaches for the nuclear modification factor $R_{AA}(p_T)$, and the LHC data. In the left plot we compare the small $x$ evolution and the energy loss approaches and find an extremely good agreement. In the right plot we compare the predictions for the approach based upon solution of the full DGLAP evolution equations to their small $x$ approximation. Note that the full $x$ DGLAP evolution method is closer in shape to the data at low $p_T<20\text{\,GeV}$ where the difference between the approaches is visible. Also note that the best value for in-medium coupling extracted from the full $x$ evolution and energy loss approaches compared to data, differs by less than~$10\%$.

\section{Conclusions}
In this talk we reviewed our results from~[\refcite{Kang:2014xsa}]. We formulated the jet quenching phenomenology in the language of DGLAP evolution equations, where the previously calculated from \SCETG medium-induced splitting kernels~[\refcite{Ovanesyan:2011kn}] are included. By solving the evolution equations analytically in the small-$x$ limit we found a deep connection between the evolution and the energy loss approaches. The evolution approach has an advantage that we can consistently implement the finite-$x$ corrections, by turning on all the four medium-induced splitting functions in the evolution equations. We found a very good agreement between the full-$x$ evolution and the energy loss approaches for the nuclear modification factor, with the extracted in-medium coupling found from comparing the two methods to data within $10\%$. The finite-$x$ corrections are important for making accurate predictions and understanding the theoretical uncertainties on the perturbative QCD side.
\section*{Acknowledgments}

I would like to thank the organizers of the 2014 QCD evolution workshop in Santa Fe, NM. Also I would like to thank my collaborators on the work reported herein: Zhong-Bo Kang, Robin Lashof-Regas, Philip Saad and Ivan Vitev. Special thanks to Ivan Vitev for his feedback on the manuscript.

\appendix

\section{Leading order $1\rightarrow 2$ splitting functions}\label{sec:appendix}
Using the results from Ref.[\refcite{Ovanesyan:2011kn}] and definitions in \eq{eq:Pdef1} and \eq{eq:Pdef2} we can unambiguously derive the functions $h_i(x,\vc{Q}_{\perp};\beta)$. We use a rewritten more compact version of the medium-induced splitting kernels, that can be found in Refs.~[\refcite{Ovanesyan:2012fr,Fickinger:2013xwa}]
\begin{eqnarray}
h_i(x,\vc{Q}_{\perp};\beta)=\vc{Q}_{\perp}^2\int\frac{d\Delta z}{\lambda_i(z)}\d^2\vc{q}_{\perp}
\frac{1}{\sigma_{\text{el}}}\frac{\d\sigma_{\text{el}}}{\d^2\vc{q}_{\perp}}\sum_{k=1}^{5}c_k^{(i)}(1-\cos\Phi_k).
\end{eqnarray}
The splitting type is described by $i$ and goes over $q\rightarrow qg$, $g\rightarrow gg$, $g\rightarrow q\bar{q}$, $q\rightarrow gq$ and $\lambda_i$ equals to gluon scattering length $\lambda_g$ for the first two splittings and to $\lambda_q$ for the last two splittings. We have defined the following two dimensional transverse vectors:
\begin{eqnarray}
\vc{A}_{\perp}=\vc{Q}_{\perp},\,\,\,\,\, \vc{B}_{\perp}=\vc{Q}_{\perp}+x\,\vc{q}_{\perp},\,
\,\,\,\,\vc{C}_{\perp}=\vc{Q}_{\perp}-(1-x)\vc{q}_{\perp},\,\,\,\,\,\vc{D}_{\perp}=\vc{Q}_{\perp}-\vc{q}_{\perp},\nonumber
\end{eqnarray}
and the following five phases $\Phi_k$ 
\begin{eqnarray}
&&\Phi_1=\Psi\vc{B}_{\perp}^2,\,\, \Phi_2=\Psi\vc{C}_{\perp}^2,\,\, \Phi_3=\Psi(\vc{C}_{\perp}^2-\vc{B}^2_{\perp}),\,\, \Phi_4=\Psi\vc{A}_{\perp}^2,\,\, \Phi_5=\Psi(\vc{A}_{\perp}^2-\vc{D}_{\perp}^2),\nonumber\\
&&\text{where}\,\, \Psi=\frac{\Delta z}{x(1-x)\,\bar{n}\mcdot p_0}.
\end{eqnarray}
The coefficients $c_k^{(i)}$ are given in the following table [\refcite{Fickinger:2013xwa}]\footnote{Note that we have changed the notation from $\alpha_k^{(i)}$ in the Appendix A of [\refcite{Fickinger:2013xwa}] to $c_k^{(i)}$ in order to avoid a possible confusion with the strong coupling $\alpha_s$.}
\begin{center}
    \begin{tabular}{ | l | l | l | p{5cm} |}
    \hline
     $k$ &$c_{k}^{(q\rightarrow qg)}$&  $c_{k}^{(g\rightarrow gg)}$ &  $c_{k}^{(g\rightarrow q\bar{q})}$ \\ \hline
    1& $\vc{b}\mcdot\left(\vc{b}-\vc{c}+\frac{\vc{a}-\vc{b}}{N_c^2}\right)$ &  $2\vc{b} \mcdot\left(\vc{b}-\vc{a}-\frac{\vc{c}-\vc{a}}{2}\right)$ & $2\vc{b} \mcdot\left(\vc{b}-\vc{a}+\frac{\vc{c}-\vc{a}}{N_c^2-1}\right)$  \\ \hline
     2&$\vc{c} \mcdot\left(2\vc{c}-\vc{a}-\vc{b}\right)$ & $2\vc{c} \mcdot\left(\vc{c}-\vc{a}-\frac{\vc{b}-\vc{a}}{2}\right)$& $2\vc{c} \mcdot\left(\vc{c}-\vc{a}+\frac{\vc{b}-\vc{a}}{N_c^2-1}\right)$ \\ \hline
  3&  $\vc{b} \mcdot\vc{c}$ & $\vc{b} \mcdot\vc{c}$ & $-2\frac{\vc{b} \,\mcdot\,\vc{c}}{N_c^2-1}$ \\ \hline
  4&   $\vc{a}\mcdot\left(\vc{d}-\vc{a}\right)$ & $\vc{a}\mcdot(\vc{d}-\vc{a})$ & $2\frac{\vc{a}\,\mcdot\,(\vc{a}-\vc{d})}{N_c^2-1}$\\ \hline
   5& $-\vc{a}\mcdot\vc{d}$ & $-\vc{a}\mcdot\vc{d}$ & $2\frac{\vc{a}\,\mcdot\,\vc{d}}{N_c^2-1}$ \\
    \hline
    \end{tabular}
\end{center}
In the table above $\vc{a}=\vc{A}_{\perp}/\vc{A}_{\perp}^2, \vc{b}=\vc{B}_{\perp}/\vc{B}_{\perp}^2, \vc{c}=\vc{C}_{\perp}/\vc{C}_{\perp}^2, \vc{d}=\vc{D}_{\perp}/\vc{D}_{\perp}^2$. Note that the coefficients for the $q\rightarrow gq$ splitting are obtained from $q\rightarrow qg$ with the substitution $x\rightarrow 1-x$.



\end{document}